\documentclass[conference]{IEEEtran}
\IEEEoverridecommandlockouts

\usepackage{cite}
\usepackage{amsmath,amssymb,amsfonts}
\usepackage{algorithmic}
\usepackage{graphicx}
\usepackage{textcomp}
\usepackage{xcolor}
\usepackage{enumerate}
\def\BibTeX{{\rm B\kern-.05em{\sc i\kern-.025em b}\kern-.08em
    T\kern-.1667em\lower.7ex\hbox{E}\kern-.125emX}}
\usepackage[capitalise, nameinlink, noabbrev, nosort]{cleveref}

\newtheorem{thm}{Theorem}
\newtheorem{defn}[thm]{Definition}
\newtheorem{lem}[thm]{Lemma}
\newtheorem{remark}[thm]{Remark}
\newtheorem{cor}[thm]{Corollary}
\newtheorem{claim}[thm]{Claim}
\newtheorem{notation}[thm]{Notation}

{\def\proof{\@IEEElegacywarn{proof}{IEEEproof}\IEEEproof}
}

\begin{document}

\title{Consistent sampling of Paley-Wiener functions on graphons
\thanks{This work was supported by the National Science Foundation Grant DMS-2408008 and Simons Travel Support for Mathematicians (to MG). HF acknowledges funding by the Deutsche Forschungsgemeinschaft (DFG, German Research Foundation)
- Project number 442047500 through the Collaborative Research Center ”Sparsity and Singular Structures” (SFB 1481).}}

\author{\IEEEauthorblockN{Hartmut F\"{u}hr}
\IEEEauthorblockA{\textit{Chair for Geometry and Analysis} \\
\textit{RWTH Aachen University}\\
Aachen, Germany \\
fuehr@mathga.rwth-aachen.de }
\and
\IEEEauthorblockN{Mahya Ghandehari}
\IEEEauthorblockA{\textit{Department of Mathematical Sciences} \\
\textit{University of Delaware}\\
Newark, USA \\
mahya@udel.edu}
}

\maketitle

\begin{abstract}
We study sampling methods for Paley-Wiener functions on graphons, thereby adapting and generalizing methods initially developed for graphs to the graphon setting. We then derive conditions under which such a sampling estimate is consistent with graphon convergence.
\end{abstract}


\section{Introduction}

A graphon $w$ can be interpreted as a probability distribution on random graphs, sampled via the \emph{$w$-random graph} process ${\mathcal G}(n,w)$, defined as follows.
Given the vertex set with labels $\{ 1,2,\dots ,n\}$, edges are formed according to $w$ in two steps. First, each vertex $i$ is assigned a value $x_i$ drawn uniformly at random from $[0,1]$. Next, for each pair of vertices with labels $i<j$ independently, an edge $\{ i,j\}$ is added with probability $w(x_i,x_j)$. 
It is known that the sequence $\{{\mathcal G}(n,w)\}_{n\in  {\mathbb N}}$  almost surely forms a convergent graph sequence, for which the limit object is the graphon $w$
(see \cite{Lovasz-Szegedy-2006}). 

In the context of graph signal processing, graphons have been proposed as a framework to develop and study signal processing techniques that are consistent across classes of similar graphs \cite{RuizChamonRibeiro21,ghandehari-etal}. In this context graph convergence provides a method of identifying similarity in graphs, and then consistency of a method can be understood as the property of being compatible with convergence to the limit object. 

Our paper can be viewed as an application of this paradigm to the problem of Shannon sampling, initially studied for graphs in \cite{Pesenson-TransAMS-2008}. We first adapt the average sampling methods from \cite{Pesenson-JFA-2021} to the graphon setting, 
and then a prove a consistency property for these methods, showing their compatibility with graphon convergence.

\section{Notations and background}

For a graph $G$, we let $V(G)$ and $E(G)$ denote its vertex set and edge set respectively.
Throughout this paper, we focus on simple graphs, i.e., undirected graphs without loops and multiple edges. 
For $n\in {\mathbb N}$, let $[n]$ denote the set $\{1,\ldots,n\}$. 
We think of ${\mathbb C}^n$ as the vector space, equipped with the inner product $\langle X,Y\rangle=\sum_{i\in [n]} X_i\overline{Y_i}$. 
We equip the interval $[0,1]$ with its Lebesgue measure, and for every measurable subset $S\subseteq [0,1]$, we denote its Lebesgue measure by $|S|$. 
For every such $S$, $L^2(S)$ denotes the vector space of square-integrable, Lebesgue measurable functions on $S$ equipped with inner product $\langle f,g\rangle_{L^2(S)}=\int_S f(x) \overline{g(x)}dx$, when $dx$ denotes the restriction of the Lebesgue measure on $S$.
%

\subsection{Convergence of graphs, graphons and $w$-random graphs}
\label{subsec:Convergence of graphs, graphons and w-random graphs}
For simple graphs $F$ and $G$, let ${\rm hom}(F, G)$ denote the number of homomorphisms of $F$ into $G$; i.e., the number of maps $V(F)\to V(G)$ that preserve edges. The homomorphism density of $F$ into $G$, defined as
$ 
 t(F,G)=\frac{{\rm hom}(F,G)}{|V(G)|^{|V(F)|}},
$
allows us to define the notion of convergent graph sequences. 
Let $\{G_n\}_{n\in {\mathbb N}}$ be a sequence of simple graphs such that $|V(G_n)| \rightarrow \infty$. 
We say that $\{G_n\}_{n\in {\mathbb N}}$ \emph{converges} if for every simple graph $F$, the numerical sequence $\{ t(F,G_n)\}_{n\in {\mathbb N}}$ is Cauchy. 
Every convergent graph sequence admits a limit that can be interpreted as a graphon.
\emph{Graphons} are measurable functions $w: [0,1]^2 \to [0,1]$ that are symmetric, 
i.e.~$w(x, y) = w(y, x)$ for almost every point $(x, y)$ in $[0, 1]^2$. 

Let ${\mathcal W}_0$ denote the set of all {graphons}, and ${\mathcal W}$ denote the (real) linear span of ${\mathcal W}_0$. 
Let  $G$ be a graph on $n$ vertices labeled $ \{ 1,2,\dots ,n\}$. 
The graph $G$ can be identified with a $\{0,1\}$-valued graphon $w_G$ as follows: split $[0, 1]$ into $n$ equal-sized intervals $\{I_i\}_{n\in [n]}$.
For $i, j\in [n]$, the graphon $w_G$ attains 1 on $I_i \times I_j$ precisely when vertices with labels $i$ and $j$ are adjacent. Note that $w_G$ depends on the labeling of the vertices of $G$, i.e., relabeling $V(G)$ may result in a different graphon.

\subsection{Cut norm and converging graph sequences} 
\label{subsec:Cut norm, cut metric and converging graph sequences}
The topology described by convergent (dense) graph sequences can be formalized by endowing ${\mathcal W}$ with the cut-norm, introduced in 
\cite{cut-norm}. For  $w \in {\mathcal W}$, the \emph{cut-norm} is defined as:
\begin{equation*}
\label{cutnorm} 
\| w \|_{\Box}= \sup_{S,T \subset [0,1]}\left|\int_{S \times T} w(x,y)\,  dxdy \right|,
\end{equation*}
where the supremum is taken over all measurable subsets $S,T$. 
That the graph sequence $\{G_n\}$ is convergent 
to $w \in {\mathcal W}_0$ is equivalent to 
the existence of suitable vertex labelings of each of the graphs $G_n$ so that we have
$\lVert w_{G_n} - w \rVert_{\square} \rightarrow 0.$
See \cite[Theorem 2.3]{BCLSV2011} for the above convergence results.

\subsection{Graphon operators} 
\label{subsec:Spectral decomposition of graphons}
For a graphon $w$, the \emph{graphon adjacency operator} $T_w$ and the \emph{graphon Laplacian operator} $L_w$ are operators on $L^2[0,1]$ defined as follows:  for  $f\in L^2[0,1]$ and a.e.~$x\in [0,1]$,
\begin{align}
T_w(f)(x)&=\int_0^1 w(x,y)f(y)\, dy, \label{eq:adjacency}\\
L_w(f)(x)&=\int_0^1 w(x,y)(f(x)-f(y))\, dy. \label{eq:Laplacian}
\end{align}
It is known that $T_w$ and $L_w$ are bounded self-adjoint operators. In addition, $T_w$ is compact and $L_w$ is positive semidefinite. 
The operator $T_w$ has a countable spectrum lying in the interval $[-1,1]$ for which 0 is the only possible accumulation point. 

%
%
%
%
%

For more details on graph limit theory, see \cite{lovasz-book}.

\section{Graphon signal sampling}
\label{sec:Graphon signal sampling}
In this section, we extend the results from \cite{Pesenson-JFA-2021} to the setting of graphons, providing analogous statements for graphons.

\begin{thm}\label{thm:general-formula}
Let $w\in {\cal W}_0$ be a graphon, and consider a partition $\{S_1,\ldots,S_k\}$ of  $[0,1]$ into measurable subsets. 
For $j\in[k]$, let $w_j$ denote the restriction of $w$ to ${S_j\times S_j}$, and $L_j$ be the associated Laplacian operator on $L^2(S_j)$ defined similar to Equation~\eqref{eq:Laplacian}. 
For each $j\in[k]$, pick $\psi_j\in L^2(S_j)$ such that $\|\psi_j\|=1$ and $\int_{S_j}\psi_j\neq 0$.
Suppose that, for each $j\in[k]$, we have the following:
\begin{enumerate}[(i)]
%
\item\label{enum-label-assump2} there exists $\delta_j>0$ such that for every $f\in L^2(S_j)$ satisfying $\int_{S_j}f=0$, we have $\|L_j^{\frac{1}{2}}f\|\geq \delta_j\|f\|$.
\end{enumerate}
Then, for every $f\in L^2[0,1]$ and for $\epsilon>0$,  we have
\begin{equation*}
\|f\|_2^2\leq (1+\epsilon)\sum_{j\in[k]}\left(\frac{|S_j|\|L_j^{\frac{1}{2}}f_j\|^2}{\delta_j^2|\int_{S_j}\psi_j|^2}+\frac{|S_j||\langle\psi_j,f_j\rangle|^2}{\epsilon|\int_{S_j}\psi_j|^2} \right).
\end{equation*}
\end{thm}
\begin{IEEEproof}
Let $f\in L^2[0,1]$ be arbitrary, and for $j\in[k]$, let $f_j\in L^2(S_j)$ denote the restriction of $f$ to $S_j$.
For each $j$, the function $\phi_j=\frac{{\mathbf 1}_{S_j}}{\sqrt{|S_j|}}\in L^2(S_j)$ is a unit eigenfunction of $L_j$  associated with eigenvalue 0. 
Condition \eqref{enum-label-assump2} implies that 0 is a simple eigenvalue of $L_j$. So, the function $\phi_j$ is an eigenvector of $L_j^{\frac12}$ associated with its simple eigenvalue 0 as well. 
\begin{claim}\label{claim:gen-PP3.2}
Let $j\in [k]$, and consider $L_j:L^2(S_j)\to L^2(S_j)$.
For every $g\in L^2(S_j)$ satisfying $\langle g,\psi_j\rangle=0$, we have 
$$\|L_j^{\frac{1}{2}}g\|\geq \delta_1|\langle\psi_j,\phi_j\rangle| \|g\|.$$
\end{claim}
{\it Proof of claim:}
It is easy to see that $L_j$ is a bounded positive semidefinite operator on $L^2(S_j)$.
Consider the closed subspace ${\cal H}_j:=\{h\in L^2(S_j):\ \langle h,\phi_j\rangle=0\}$ of $L^2(S_j)$, and let $P_j:L^2(S_j)\to {\cal H}_j$  denote the associated orthogonal projection.

Let $g$ be any element of $L^2(S_j)$ satisfying $\langle g,\psi_j\rangle=0$,
and write $g=g_1+g_2$, where $g_1=\langle g,\phi_j\rangle \phi_j$ and $g_2=P_jg$.
Since $\langle g, \psi_j\rangle=0$, we have $\langle g_1,\psi_j\rangle=-\langle g_2, \psi_j\rangle$, which can be written as
$\langle g,\phi_j\rangle \langle \phi_j,\psi_j\rangle=-\langle g-g_1, \psi_j\rangle.$
So, we have 
\begin{align*}
|\langle \psi_j,\phi_j\rangle|^2 \|g\|^2&=|\langle \psi_j,\phi_j\rangle|^2 (|\langle g,\phi_j\rangle|^2+\|g-g_{1}\|^2)\\
&=|\langle g-g_{1}, \psi_j\rangle|^2+ |\langle \psi_j,\phi_j\rangle|^2 \|g-g_{1}\|^2\\
&=  |\langle g-g_{1}, P_j\psi_j\rangle|^2+ |\langle \psi_j,\phi_j\rangle|^2 \|g-g_{1}\|^2,
\end{align*}
where in the last equation, we used the fact that 
$P_j(g-g_{1})=g-g_{1}$. Applying the Cauchy--Schwartz inequality, we get
\begin{align*}
|\langle \psi_j,\phi_j\rangle|^2 \|g\|^2
&\leq \|g-g_{1}\|^2\|P_j\psi_j\|^2+ |\langle \psi_j,\phi_j\rangle|^2 \|g-g_{1}\|^2\\
&=\|g-g_{1}\|^2\|\psi_j\|^2.
\end{align*}
Noting that $g-g_1=P_jg$ and $\|\psi_j\|=1$, we get 
\begin{equation}\label{claim-eq}
    |\langle \psi_j,\phi_j\rangle| \|g\|\leq \|P_jg\|.
\end{equation}
Since $\phi_j$ is an eigenvector of $L_j^{\frac{1}{2}}$ associated with eigenvalue 0, we have $L_j^{\frac{1}{2}}\phi_j=0$, and thus,  $L_j^{\frac{1}{2}}g=L_j^{\frac{1}{2}}(g_1+P_jg)=L_j^{\frac{1}{2}}P_jg$.  So, by condition \eqref{enum-label-assump2} of \Cref{thm:general-formula} and \eqref{claim-eq}, we have
\begin{align*}
\|L_j^{\frac{1}{2}}g\|=\|L_j^{\frac{1}{2}}P_jg\|\geq \delta_j\|P_jg\|\geq \delta_j|\langle \psi_j,\phi_j\rangle| \|g\|.
\end{align*}
This finishes the proof of the claim.

Applying Claim \ref{claim:gen-PP3.2} to $f_j-\frac{\langle f_j,\psi_j\rangle}{\langle \phi_j,\psi_j\rangle}\phi_j$, we get 
\begin{equation}\label{eq-part1}
\|L_j^{\frac{1}{2}}(f_j-\frac{\langle f_j,\psi_j\rangle}{\langle \phi_j,\psi_j\rangle}\phi_j)\|\geq \delta_j|\langle \psi_j, \phi_j\rangle|\|f_j-\frac{\langle f_j,\psi_j\rangle}{\langle \phi_j,\psi_j\rangle}\phi_j\|.
\end{equation}
Let $\epsilon>0$ be arbitrary, and note that for any nonnegative numbers $a,b$, we have
$(\sqrt{\epsilon}a-\frac{1}{\sqrt{\epsilon}}b)^2\geq 0$. This inequality can be equivalently written as $(a+b)^2\leq (1+\epsilon)a^2+\frac{1+\epsilon}{\epsilon}b^2$. Combining this fact with the triangle inequality, we get
\begin{equation*}
    \|f_j\|^2\leq (1+\epsilon)\left\|f_j-\frac{\langle f_j,\psi_j\rangle\phi_j}{\langle \phi_j,\psi_j\rangle}\right\|^2+\frac{1+\epsilon}{\epsilon}\left\|\frac{\langle f_j,\psi_j\rangle\phi_j}{\langle \phi_j,\psi_j\rangle}\right\|^2.
\end{equation*}
The inequality above, combined with \eqref{eq-part1} and the fact that $L_j^{\frac{1}{2}}\phi_j=0$, implies that
\begin{align*}
    \|f_j\|^2
    &\leq(1+\epsilon)\frac{\|L_j^{\frac{1}{2}}f_j\|^2}{\delta_j^2|\langle\psi_j,\phi_j\rangle|^2}+\frac{1+\epsilon}{\epsilon}\left|\frac{\langle f_j,\psi_j\rangle}{\langle \phi_j,\psi_j\rangle}\right|^2\nonumber\\
    &=(1+\epsilon)\frac{|S_j|\|L_j^{\frac{1}{2}}f_j\|^2}{\delta_j^2|\int_{S_j}\psi_j|^2}+\frac{1+\epsilon}{\epsilon}\frac{|S_j|}{|\int_{S_j}\psi_j|^2} |\langle\psi_j,f_j\rangle|^2,
\end{align*}
which finishes the proof, since $\|f\|^2=\sum_{j=1}^k\|f_j\|^2$. 
\end{IEEEproof}

\begin{remark}
We say $w_j:S_j\times S_j\to [0,1]$ is \emph{connected}, if for every measurable subset $S\subseteq S_j$ with $0<|S|<|S_j|$, 
$\int_{S\times (S_j\setminus S)}w(x,y)\, dx\, dy>0.$
Condition \eqref{enum-label-assump2} implies that 0 is a simple eigenvalue of $L_j$.
We claim that $0$ is a simple eigenvalue of $L_j$ if $w_j$ is connected. 
Clearly, ${\mathbf 1}_{S_j}$ is a 0-eigenvector of $L_j$. 
Suppose that $0\neq f\in L^2(S_j)$ is another eigenvector of $L_j$ associated with 0 such that $f\perp{\mathbf 1}_{S_j}$.
Let $E:=f^{-1}(0,\infty)$, and observe that $0<|E|<|S_j|$.
Clearly, $S_j\setminus E:=f^{-1}(-\infty,0]$. 
Next, we have
\begin{eqnarray*}
   0&=&\langle L_jf,f\rangle=\frac{1}{2}\iint_{S_j\times S_j} w(x,y)(f(x)-f(y))^2\\
&\geq& \frac{1}{2}\iint_{E\times (S_j\setminus E)} w(x,y)(f(x)-f(y))^2.
\end{eqnarray*}
Since $f(x)-f(y)>0$ for $(x,y)\in E\times (S_j\setminus E)$,
the last integral in the above expression is zero precisely when $w|_{E\times (S_j\setminus E)}\equiv 0$ a.e.;
this is a contradiction as $w$ is connected.
\end{remark}
\begin{notation}\label{notation1}
Let $j\in [k]$. We define the following notations.
\begin{equation*}
    \theta_j:=\frac{|S_j|}{|\int_{S_j}\psi_j|^2},\
    \theta:=\max_{j\in[k]} \theta_j,\
    \delta:=\min_{j\in [k]} \delta_j.
\end{equation*}
\end{notation}

\begin{cor}\label{cor:general-formula}
With notations and assumptions from \Cref{thm:general-formula} and Notation \ref{notation1}, 
for every $f\in L^2[0,1]$ and for $\epsilon>0$, 
\begin{equation*}
\|f\|^2\leq \frac{(1+\epsilon)\theta}{\delta^2}\|L_w^{\frac{1}{2}}f\|^2+\frac{1+\epsilon}{\epsilon}\theta \sum_{j=1}^k|\langle\psi_j,f\rangle|^2.
\end{equation*}
\end{cor}
\begin{IEEEproof}
From \Cref{thm:general-formula}, the expression 
$$(1+\epsilon)\sum_{j=1}^k\frac{|S_j|\|L_j^{\frac{1}{2}}f_j\|^2}{\delta_j^2|\int_{S_j}\psi_j|^2}+\frac{1+\epsilon}{\epsilon}\sum_{j=1}^k\frac{|S_j|}{|\int_{S_j}\psi_j|^2} |\langle\psi_j,f_j\rangle|^2$$
is an upper bound for $\|f\|^2$. Using Notation \ref{notation1}, we have
\begin{align*}
\|f\|^2
&\leq\frac{(1+\epsilon)\theta}{\delta^2}\sum_{j=1}^k\|L_j^{\frac{1}{2}}f_j\|^2+\frac{(1+\epsilon)\theta}{\epsilon}\sum_{j=1}^k |\langle\psi_j,f_j\rangle|^2\\
&=\frac{(1+\epsilon)\theta}{\delta^2}\sum_{j=1}^k\|L_j^{\frac{1}{2}}f_j\|^2+\frac{(1+\epsilon)\theta}{\epsilon}\sum_{j=1}^k |\langle\psi_j,f\rangle|^2,
\end{align*}
where in the last equality we used the fact that each $\psi_j$ is supported in $S_j$.

To finish the proof, we only need to show that
$\sum_{j=1}^k\|L_j^{\frac{1}{2}}f_j\|^2\leq \|L_w^{\frac{1}{2}}f\|^2$. To see this, note that 
\begin{align*}
    \|L_j^{\frac{1}{2}}f_j\|^2&=\frac{1}{2}\iint_{S_j\times S_j} w_j(x,y)(f_j(x)-f_j(y))^2\, dxdy\\
    &=\frac{1}{2}\iint_{S_j\times S_j} w(x,y)(f(x)-f(y))^2\, dxdy,
\end{align*}
which implies that 
$$\sum_{j=1}^k\|L_j^{\frac{1}{2}}f_j\|^2\leq \frac{1}{2}\iint_{[0,1]^2} w(x,y)(f(x)-f(y))^2=\|L_w^{\frac{1}{2}}f\|_2^2.$$
\end{IEEEproof}
\begin{defn}
For $\tau_j>0$, we define the set 
$$\chi_j(\tau_j):=\left\{f\in L^2(S_j):\ \|L_j^{\frac{1}{2}}f\|_2\leq \tau_j\|f\|_2\right\}.$$
\end{defn}

\begin{cor}\label{cor:appl-graphon}
Suppose all assumptions of \Cref{thm:general-formula} hold. Let $0\leq \sigma<1$. 
For every $j\in [k]$, choose $\tau_j>0$ satisfying 
$\frac{\theta_j}{\delta_j^2}\tau_j^2\leq \sigma$. For every $f\in L^2[0,1]$ satisfying 
$f|_{S_j}:=f_j\in \chi_j(\tau_j)$, we have
$$\frac{(1-(1+\epsilon)\sigma)\epsilon}{(1+\epsilon)\theta}\|f\|^2\leq \sum_{j=1}^k|\langle f,\psi_j\rangle|^2\leq \|f\|^2,$$
where $\epsilon>0$ is chosen such that $(1+\epsilon)\sigma<1.$
\end{cor}
\begin{IEEEproof}
The inequality $\sum_{j=1}^k|\langle f,\psi_j\rangle|^2\leq \|f\|^2$ follows from the fact that $\{\psi_j: j\in [k]\}$ is an orthonormal set.
For the other inequality, using \Cref{thm:general-formula} and Notation \ref{notation1}, we have
\begin{align*}
\|f\|^2&\leq(1+\epsilon)\sum_{j=1}^k\frac{\theta_j\tau_j^2\|f_j\|^2}{\delta_j^2}+\frac{1+\epsilon}{\epsilon}\sum_{j=1}^k\theta_j |\langle\psi_j,f_j\rangle|^2\\
 &\leq(1+\epsilon)\sigma\sum_{j=1}^k\|f_j\|^2+\frac{1+\epsilon}{\epsilon}\theta\sum_{j=1}^k |\langle\psi_j,f_j\rangle|^2\\
 &=(1+\epsilon)\sigma\|f\|^2+\frac{1+\epsilon}{\epsilon}\theta\sum_{j=1}^k |\langle\psi_j,f_j\rangle|^2,
\end{align*}
where the first inequality follows from $f_j\in \chi_j(\tau_j)$. So,
$$\frac{(1-(1+\epsilon)\sigma)\epsilon}{(1+\epsilon)\theta}\|f\|^2\leq \sum_{j=1}^k|\langle f,\psi_j\rangle|^2.$$
\end{IEEEproof}

\begin{defn}
For $\gamma> 0$, define the spectral projection $P_\gamma={\mathbf 1}_{[0,\gamma]}(L_w)$ in the sense of functional calculus. 
The \emph{Paley-Wiener} space associated with the Laplacian operator $L_w$ is defined as the image of the above  projection, and is denoted by $PW_\gamma(w)$, i.e.,
$PW_\gamma(w) =  P_\gamma(L^2[0,1]).$
\end{defn}

\begin{cor}\label{cor-PW-bound}
With terminology from Notation \ref{notation1}, let $\gamma>0$ be such that  $\gamma<\frac{\delta^2}{\theta}$. 
With assumptions from \Cref{thm:general-formula}, for every $f\in PW_\gamma(w)$ we have
$$\frac{(\delta-\sqrt{\theta\gamma})^2}{\theta\delta^2}\|f\|^2\leq \sum_{j=1}^k|\langle f,\psi_j\rangle|^2\leq \|f\|^2.$$
\end{cor}

\begin{IEEEproof}
By Corollary \ref{cor:general-formula}, for every $f\in PW_\gamma(w)$ and $\epsilon>0$, 
\begin{eqnarray*}
\|f\|^2&\leq& \frac{(1+\epsilon)\theta}{\delta^2}\|L_w^{\frac{1}{2}}f\|^2+\frac{1+\epsilon}{\epsilon}\theta \sum_{j=1}^k|\langle\psi_j,f\rangle|^2\\
&\leq& \frac{(1+\epsilon)\theta}{\delta^2}\gamma\|f\|^2+\frac{1+\epsilon}{\epsilon}\theta \sum_{j=1}^k|\langle\psi_j,f\rangle|^2,
\end{eqnarray*}
where in the second inequality we used $\|L_w^{\frac{1}{2}}f\|\leq \sqrt{\gamma}\|f\|$ for $f\in PW_\gamma(w)$.
So, for $\epsilon\in (0,\frac{\delta^2}{\theta\gamma}-1)$, we get the following:
\begin{equation}\label{eq:to-opt}
\frac{(1-\frac{(1+\epsilon)\theta}{\delta^2}\gamma){\epsilon}}{(1+\epsilon)\theta}\|f\|^2\leq  \sum_{j=1}^k|\langle\psi_j,f\rangle|^2.
\end{equation}
To optimize inequality \eqref{eq:to-opt}, we observe that $\epsilon=\frac{\delta}{\sqrt{\theta\gamma}}-1$ lies  within the appropriate interval $(0,\frac{\delta^2}{\theta\gamma}-1)$ and maximizes the function $f(\epsilon)=\frac{(1-\frac{(1+\epsilon)\theta\gamma}{\delta^2}){\epsilon}}{(1+\epsilon)\theta}$. Plugging $\epsilon=\frac{\delta}{\sqrt{\theta\gamma}}-1$ in the left hand side of \eqref{eq:to-opt} finishes the proof. 
\end{IEEEproof}

\section{Sampling from converging graph sequences}
\label{sec:convergence-results}
Let $w$ and $w_n$, for $n\in {\mathbb N}$, denote graphons. 
\begin{defn}\label{def:degree-mult}
    The \emph{degree function} of a graphon $w$ is defined as follows. For almost every $x\in[0,1]$,
    $$d_w:[0,1]\to [0,1], \ d_w(x):=\int_0^1w(x,y)\, dy.$$ 
For a graphon $w$ with degree function $d_w$, the associated \emph{multiplication operator} is defined as 
\begin{equation*}
    M_w:L^2[0,1]\to L^2[0,1], \ M_wf(x)=d_w(x)f(x), \ \mbox{ a. e. }
\end{equation*}
\end{defn}
\begin{lem}\label{lem:M-WOT}
Suppose $\lim_{n\to \infty}\|w_n-w\|_\Box= 0$.
Then $M_{w_n}\to M_w$ in the weak operator topology (WOT).
As a consequence, $L_{w_n}\to L_w$ in WOT as well. 
\end{lem}
\begin{IEEEproof}
For a measurable subset $S$ of $[0,1]$, we have
$\lim_{n\to \infty}\iint_{S\times [0,1]} (w_n(x,y)-w(x,y))\, dx\, dy=0,$ as $w_n\to w$ in cut-norm.
So, for every measurable subset $S\subseteq [0,1]$, 
\begin{equation}
\label{eq:lim-d}
    \lim_{n\to \infty}\int_{0}^1(d_{w_n}(x)-d(x)){\mathbf 1}_S(x)\, dx=0.
\end{equation}
Using the fact that step functions are dense in $L^1[0,1]$ and $\|d_{w_n}-d\|_\infty\leq 2$, and applying H\"{o}lder's inequality, we can 
extend \eqref{eq:lim-d} to obtain the following:
\begin{equation}\label{eq:dn-conv}
    \lim_{n\to \infty}\int_0^1(d_{w_n}(x)-d(x))h(x)\, dx=0, \ \forall\ h\in L^1[0,1].
\end{equation}
Now, let $f,g\in L^2[0,1]$ be arbitrary. Since $f\overline{g}\in L^1[0,1]$, applying \eqref{eq:dn-conv}, we get
$\lim_{n\to\infty}\langle (M_{w_n}-M_w)f,g\rangle= 0.$

To show that $L_{w_n}\to L_w$ in WOT, we only need to verify the convergence $T_{w_n}\to T_w$ in WOT, given that $L_w=M_w-T_w$ for any graphon $w$. 
Now, applying  \cite[Equation 4.4 and Lemma E.6]{janson}, we observe that $\{T_{w_n}\}_{n\in {\mathbb N}}$ converges to $T_w$ in the operator norm. This finishes the proof, as convergence in operator norm implies convergence in WOT.
\end{IEEEproof}

Using Lemma \ref{lem:M-WOT}, if $\lim_{n\to \infty}\|w_n-w\|_\Box= 0$ then for every $f\in L^2[0,1]$ we have $\lim_{n\to \infty} \|L_{w_n}^{\frac12}f\|^2= \|L_w^{\frac12}f\|^2$. 
Under the conditions of \Cref{thm:general-formula}, approximating $\|L_{w_n}^{\frac12}f\|^2$ by $\|L_{w}^{\frac12}f\|^2$ from below and appealing to \Cref{thm:general-formula}, we get that for every $f\in L^2[0,1]$, there exists a large enough index $N$ such that for every $n\geq N$, if $f\in PW_\gamma(w_n)$ then
$$\frac{(\delta-\sqrt{\theta\gamma})^2}{2\theta\delta^2}\|f\|^2\leq \sum_{j=1}^k|\langle f,\psi_j\rangle|^2\leq \|f\|^2.$$

To prove a more friendly robustness result in sampling, we need a stronger convergence, namely the operator norm convergence, of the graphon Laplacian operators. 
Adding extra assumptions on the sequence of degree functions, we show that the sampling rate for a given $f$ belonging to the Paley-Wiener space of $w_n$, for large enough $n$, is independent of $n$. This can be interpreted as robustness in sampling.
\begin{thm}\label{thm:convergence}
Suppose $\lim_{n\to \infty}\|w_n-w\|_\Box= 0$.
Suppose, in addition, that $\lim_{n\to \infty}\|d_n-d\|_\infty= 0$.
Let $\{S_j\}_{j\in [k]}$ and $\psi_j\in L^2(S_j)$ be as in \Cref{thm:general-formula}, and let $\theta, \delta$ denote the constants associated to $w, \{ S_j \}, \{ \psi_j \}$ according to Notation~\ref{notation1}.
Let $\gamma>0$ such that  $\gamma<\frac{\delta^2}{\theta}$. 
There exists $N\in {\mathbb N}$, such that for all $n\geq N$, if $f\in PW_\gamma(w_n)$ then
$$\frac{(\delta-\sqrt{\theta\gamma})^2}{2\theta\delta^2}\|f\|^2\leq \sum_{j=1}^k|\langle f,\psi_j\rangle|^2\leq \|f\|^2.$$
\end{thm}
\begin{IEEEproof}
It is well-known that the norm of a multiplication operator on $L^2[0,1]$ is given  by the $L^\infty$-norm of the multiplying function. So, 
$\|M_{w_n}-M_w\|_{\rm{opr}}=\|d_n-d\|_\infty\to 0$ as $n$ tends to infinity. Consequently, 
$\lim_{n\to \infty}\|L_{w_n}-L_w\|_{\rm{opr}}=0$. 
The following argument can be understood as a perturbed version of the proof of Corollary \ref{cor-PW-bound}.

We fix $\epsilon\in (0,\frac{\delta^2}{\theta\gamma}-1)$ and $\epsilon'>0$; both will be specified further at the end of the proof. Choose an index $N$ so that $\|L_{w_n}-L_w\|_{\rm{opr}}<\epsilon'$
for all $n\geq N$. This then entails for every $f\in L^2[0,1]$ and $n\geq N$, that
\begin{eqnarray*}
   \left|\|L_{w_n}^{\frac12}f\|^2-\|L_{w}^{\frac12}f\|^2\right|= \left|\langle L_{w_n}f,f\rangle-\langle L_{w}f,f\rangle\right|\leq \epsilon'\|f\|^2~.
\end{eqnarray*}
Now assume that $n \ge N$ and $f \in PW_\gamma(w_n)$. Using Corollary \ref{cor:general-formula} together with $\|L_{w}^{\frac{1}{2}} f\|^2 \le \|L_{w_n}^{\frac{1}{2}} f\|^2 + \epsilon' \| f \|^2$ then provides the estimate 
\begin{align*}
\|f\|^2 
&\leq   \frac{(1+\epsilon)\theta}{\delta^2}\gamma\|f\|^2+\frac{1+\epsilon}{\epsilon}\theta \sum_{j=1}^k|\langle\psi_j,f\rangle|^2 \\
 & +   \epsilon' \frac{(1+\epsilon)\theta}{\delta^2} \|f\|^2~. 
\end{align*}
Now fix $\epsilon=\frac{\delta}{\sqrt{\theta\gamma}}-1$, as in the proof of Corollary \ref{cor-PW-bound}. We then obtain the estimate
\[
\frac{(\delta-\sqrt{\theta\gamma})^2}{\theta\delta^2}\|f\|^2\leq \sum_{j=1}^k|\langle f,\psi_j\rangle|^2 + \epsilon' \frac{\epsilon}{\delta^2} \| f \|^2~. 
\] 
Picking
$\epsilon' \le \frac{ (\delta - \sqrt{\theta \gamma})^2}{2 \epsilon \theta}$ provides the desired conclusion. 
\end{IEEEproof}

\begin{remark}
Theorem \ref{thm:convergence} can be understood as a sampling theorem that is consistent in the sense discussed in the introduction: Both the sampling functionals and the constants are determined from the limit object $w$, but they give rise to sampling estimates that are uniform for all approximants $w_n$ which are sufficiently close to the limit object. 

While these aspects of the theorem are rather satisfactory, there is reason to believe that it can be improved substantially. Most importantly, the assumption that the degree functions converge uniformly is rather strong. Note that this property does not generally follow from cut-norm convergence.

That said, there are some easily-identified settings in which uniform convergence actually holds. As a class of examples, consider a sequence $\{w_n\}_{n \in \mathbb{N}}$ of step graphons, where $w_n$ is obtained by averaging a fixed graphon $w$ over squares of size $\frac{1}{n} \times \frac{1}{n}$. 
Then $w_n$ converge to $w$ in cut-norm, and the associated degree functions $d_{w_n}$ can be obtained directly from the degree function $d_w$, by averaging over intervals of length $\frac{1}{n}$. 
It is easy to check that \Cref{thm:convergence} applies to this sequence of graphons as soon as $d_w$ is the uniform limit of such averages. The class of functions for which this convergence statement holds is fairly large, containing (for example) piecewise continuous functions possessing one-sided limits at each point. 

The question of extending the theorem to allow weaker convergence assumptions is the subject of ongoing research. Another interesting and currently open question concerns the systematic construction of the partitions $\{S_1,\ldots,S_k \}$ that are needed for the approach, ideally with some control over the associated constants entering the sampling estimate.  
\end{remark}

\section*{Acknowledgment}
The authors thank the Department of Mathematical Sciences at University of Delaware and the Department of Mathematics at RWTH Aachen for funding their collaborative visit in 2024. 

\end{document}